\documentclass[twocolumn,showpacs,preprintnumbers,amsmath,amssymb]{revtex4}
\usepackage{color}
\usepackage{amsmath}
\usepackage{makeidx}
\usepackage{graphicx}
\usepackage{amsfonts}
\usepackage{amssymb}
\usepackage[latin1]{inputenc}
\usepackage[T1]{fontenc}

\newcommand{\beq}{\begin{equation}}
\newcommand{\eeq}{\end{equation}}
\newcommand{\beqr}{\begin{eqnarray}}
\newcommand{\eeqr}{\end{eqnarray}}

\begin{document}

\title{Conditional squeezing of an atomic alignment}
\author{J. Cviklinski$^1$, A. Dantan$^2$, J. Ortalo$^1$ and M. Pinard$^1$}
\affiliation{$^1$ Laboratoire Kastler Brossel, Université Pierre et
Marie Curie, 4 place Jussieu, F75252 Paris Cedex 05, France
\\$^2$ QUANTOP, Danish National Research Foundation
Center for Quantum Optics, Department of Physics and Astronomy,
University of Aarhus, DK-8000 \AA rhus C., Denmark}

\date{\today}

\begin{abstract}
We investigate the possibility to perform Quantum Non Demolition
measurements of the collective alignment of an atomic ensemble in
the case of a $F\geq 1$ spin. We compare the case of purely
vectorial and purely tensorial Hamiltonians and show how to achieve
conditional squeezing or entanglement of atomic alignment
components.
\end{abstract}

\pacs{03.67.Mn,34.80.Qb,42.50.Ct,42.50.Dv}

\maketitle

\section{Introduction}

The reduction in the quantum fluctuations of an atomic ensemble
angular momentum has recently received much attention in connection
with quantum information, high-sensitivity frequency measurements
and high-precision magnetometry. Such spin-squeezed atomic states
may be obtained via non-linear interaction processes between an
ensemble of $\Lambda$-atoms and cavity fields
\cite{dantanSelf,dantanCPT} or by direct mapping of a squeezed state
of light onto the ground state atomic spin
\cite{kozhekin,dantanEIT,dantanPRA05,polziktensbis}. Another
approach consists in probing the atomic angular momentum in a
Quantum Non Demolition (QND) manner in order to reduce the quantum
fluctuations of one of its components below the standard quantum
noise \cite{kuzmichprl,polzikX}. The atomic squeezing is then
conditioned on the QND measurement result and can be actively fed
back to the atomic angular momentum using a magnetic field
\cite{wiseman}. So far, these protocols have been implemented with
cesium atoms \cite{mabusci,polzikint}. Since the angular momentum is
greater than 1/2 a full description of the atomic state not only
requires to take into account the three components of the angular
momentum, but also the higher order tensorial components. This means
that one has to add to the simplified effective QND Hamiltonian
$F_zS_z$ \cite{mabusci,polzikint} three of the five components of
the atomic alignment, which in general will perturb the measurement
of the orientation $F_z$. It is however possible to choose the
atomic detuning with the excited states such that their contribution
is zero or negligible \cite{kuzmichpra,polziktens,deechaniz}.

The goal of the present paper is to investigate high-angular
momentum atom situations in which the Hamiltonian is not purely
vectorial and show how it is actually possible to realize QND
measurements of the atomic alignment components. Such measurements
may then allow for squeezing not only the quantum fluctuations of
the atomic orientation, but also those of the alignment, which are
involved in several atom/light quantum interface protocols
\cite{DLCZ,dantanCPT}. For instance, by achieving conditional
squeezing of the alignment of an atomic ensemble combined by single
atomic excitation retrievals using the ``DLCZ" protocol
\cite{DLCZ,memKuzmich}, it is possible in principle to produce
exotic atomic states with a non-Gaussian Wigner functions, in a way
similar to non-Gaussian optical states
\cite{grangierchaton,polzikchaton}. In addition to being a tool for
atomic quantum noise studies, controlling the fluctuations of the
atomic alignment may be of interest for improving the precision of
magnetometers \cite{magneto,mabusci,comspin}. In order to draw
simple conclusions we shall limit ourselves to a first order linear
atom-field interaction in the optical pumping regime, but we note
that interesting possibilities may also be offered by
orientation/alignment conversion \cite{oriali} and non-linear
selective addressing of high-rank atomic polarization
moments~\cite{selecttens}.

In Sec.~\ref{hamiltonien} we give the effective Hamiltonian and
derive the atom-field evolution equations. After reviewing in
Sec.~\ref{vectoriel} the well-known vectorial Hamiltonian situation
leading to QND squeezing of the orientation, we examine in
Sec.~\ref{tensoriel} the purely tensorial Hamiltonian situation. We
highlight the differences with the vectorial situation and show how
QND measurements of the alignment can be performed, leading to
conditional squeezing or entanglement of the atomic components. The
effect of spontaneous emission losses on the obtainable squeezing
and the experimental feasibility are discussed in Sec.~\ref{noise}
in the case of rubidium atoms.

\section{Hamiltonian and evolution of the system}\label{hamiltonien}

We consider an optical field propagating along $z$ which interacts
with an $N$-atom ensemble in the low saturation regime and consider
slow processes as compared to the evolution of the excited state
populations and optical coherences, which can be adiabatically
eliminated. In this case, the effective Hamiltonian describing the
atom-light interaction can be written as
\cite{happer,kuzmichpra,polziktens}
\beqr \nonumber H_{int} &=& \sum_{F'} \hbar \frac{\sigma_{F'}}{2A}
\frac{\Gamma/2}{\Delta_{F'}+ i \Gamma/2} \int^L_0 dz \big\{
    \frac{ \alpha_V^{F'}}{2}F_z\left(z,t\right)S_z
    \left(z,t\right)\\\nonumber &&
    - \frac{\alpha^T_{F'}}{2(F+1)}\left[
        F^2_z\left(z,t\right) - \frac{F(F+1)}{3}
        \right]S_0\left(z,t\right)\\\nonumber &&
    +\frac{\alpha_T^{F'}}{2(F+1)}\left(
        F^2_x - F^2_y
    \right)\left(z,t\right)S_x\left(z,t\right)\\ &&
   +\frac{\alpha_T^{F'}}{2(F+1)}\left(
        F_x F_y + F_y F_x
    \right)\left(z,t\right)S_y\left(z,t\right) \big\}\label{eq:Hint} \eeqr
F (resp. F') is the total angular momentum of the ground state
(resp. of one of the excited states), and its cartesian components
are denoted by $F_{x,y,z}$. $\sigma_{F'}$ is the resonant
cross-section of the $F\rightarrow F'$ transition, and $\Delta_{F'}$
is the probe one-photon detuning with respect to this transition
($>0$ if blue detuned). $A$ is the field cross-section and $L$ the
length of the $N$-atom medium. The vectorial and tensorial
polarisabilities are denoted by $\alpha_V^{F'}$ and $\alpha_T^{F'}$
and their exact form, given in Refs.~\cite{happer,polziktens}, is
reminded in Appendix \ref{alpha}. The definition for the Stokes
operators used throughout the paper is \beqr
S_x &=& a^{\dagger}_x a_x - a^{\dagger}_y a_y\\
S_y &=& a^{\dagger}_x a_y + a^{\dagger}_y a_x\\
S_z &=& i \left(
    a^{\dagger}_y a_x - a^{\dagger}_x a_y
    \right)
\\
S_0 &=& a^{\dagger}_x a_x + a^{\dagger}_y a_y \eeqr where the field
$a$ with frequency $\omega$ is defined by $E =\mathcal{E}_0 (a +
a^{\dagger})$ and $\mathcal{E}_0 = \sqrt{\hbar \omega/2 \epsilon_0 A
c}$.

To simplify the discussion and relate it to the experimental
situation which will be considered in Sec.~\ref{noise}, we assume in
the following an $F=1$ total ground state spin, but the physical
conclusions would actually remain the same for a higher angular
momentum. The irreducible tensor operators $T^k_q$ for $F=1$ are
given by~\cite{omont} \beqr
T^1_0 &=& F_z / \sqrt{2} \\
T^1_{\pm}&=& \pm F_{\pm} / 2\\
T^2_{0} &=& (3F^2_{z}-2)/\sqrt{6}\\
T^2_{\pm 1} &=& \pm (F_zF_{\pm}+F_{\pm}F_z)/2\\
T^2_{\pm 2} &=& (F^2_{\pm})/2 \eeqr with $F_{\pm}= F_x \pm i F_y$.
In this case, the Hamiltonian reads \beqr\nonumber H_{int} &=&
\sum_{F'} \hbar \frac{\sigma_{F'}}{2A} \frac{\Gamma/2}{\Delta_{F'}+
i \Gamma/2} \int^L_0 dz \big\{
    \frac{\alpha_V^{F'} }{\sqrt{2}} T^1_0\left(z,t\right)S_z\left(z,t\right)\\\nonumber &&- \frac{\alpha_T^{F'}
    }{2\sqrt{6}}T^2_0\left(z,t\right)S_0\left(z,t\right)\\\nonumber &&
    +\frac{\alpha_T^{F'}}{4}(T^2_2+T^2_{-2})\left(z,t\right)
    S_x\left(z,t\right)\\ &&
   +\frac{\alpha_T^{F'} }{4i}(T^2_2-T^2_{-2})\left(z,t\right)S_y\left(z,t\right)
\big\} \eeqr

The anti-hermitic terms in the Hamiltonian of Eq.~(\ref{eq:Hint})
are due to optical pumping. For an off-resonant interaction, these
anti-hermitic terms may be neglected, although their contribution
should be considered carefully when it comes to optimizing the
squeezing as it will be shown in Sec.~\ref{noise}. If these terms
are neglected the evolution of the atomic operators is simply given
by
$\frac{d}{dt}\hat{A}=\frac{1}{i\hbar}\left[\hat{A},H_{int}\right]$,
which yields
\begin{widetext}\begin{equation} \frac{d}{dt}
    \left[
        \begin{array}{c}
         T^1_0 \\
         \left(T^2_2 + T^2_{-2}\right)/ \sqrt{2} \\
         \left(T^2_2 - T^2_{-2}\right)/(i\sqrt{2})
        \end{array}
        \right]
= \sum_{F'} \frac{\sigma_{F'}\Gamma}{4A\Delta_{F'}}
     \left[
    \begin{array}{ccc}
      0 & - \alpha_T^{F'} S_y / 2 & \alpha_T^{F'} S_x / 2\\
      \alpha_T^{F'} S_y / 2 & 0 & - \alpha_V^{F'} S_z\\
      - \alpha_T^{F'} S_x / 2 & \alpha_V^{F'} S_z & 0 \\
    \end{array}
    \right]
    \left[
        \begin{array}{c}
         T^1_0 \\
         \left(T^2_2 + T^2_{-2}\right)/\sqrt{2} \\
         \left(T^2_2 - T^2_{-2}\right)/(i\sqrt{2})
        \end{array}
        \right]
        \label{evoat}
\end{equation} We limited ourselves to this set of three operators, since it is a closed system
under $\hat{H}_{int}$ and allows for conditional squeezing of $T^2_2
+ T^2_{-2}$ or $T^2_2 + T^2_{-2}$ as we will show later. The terms
$\propto \alpha_V$ in Eq.~(\ref{evoat}) correspond to light-shifts,
and the ones $\propto \alpha_T$ to Raman processes involving
coherences between sublevels with $\left|\Delta
\mathrm{m}_F\right|=2$. Under the slowly varying envelope and
paraxial approximations \cite{dantanPRA05}, the field evolution
equations read
\begin{eqnarray}
     \left(\frac{\partial}{\partial z}+\frac{1}{c}\frac{\partial}{\partial t}\right)
    \left[
        \begin{array}{c}
         S_x \\
         S_y \\
         S_z
        \end{array}
        \right]
&=&
    \sum_{F'} \frac{\sigma_{F'}\Gamma}{4A\Delta_{F'}}
     \left[
    \begin{array}{ccc}
      0 & - \sqrt{2} \alpha_V^{F'} T^1_0 & \frac{\alpha_T^{F'}}{\sqrt{2}} \frac{T^2_2 - T^2_{-2}}{i\sqrt{2}} \\
      \sqrt{2} \alpha_V^{F'} T^1_0 & 0 & -  \frac{\alpha_T^{F'}}{\sqrt{2}} \frac{T^2_2 + T^2_{-2}}{\sqrt{2}}\\
      - \frac{\alpha_T^{F'}}{\sqrt{2}} \frac{T^2_2 - T^2_{-2}}{i\sqrt{2}}
      & \frac{\alpha_T^{F'}}{\sqrt{2}} \frac{T^2_2 + T^2_{-2}}{\sqrt{2}} & 0 \\
    \end{array}
    \right]
    \left[
        \begin{array}{c}
         S_x \\
         S_y \\
         S_z
        \end{array}
        \right]
        \label{inoutphot}
\end{eqnarray}\end{widetext}
The terms $\propto \alpha_V$ in Eq.~(\ref{inoutphot}) correspond to
the well known Faraday rotation. In the following, we will consider
the Stokes operators \textit{before} ($in$) and \textit{after}
($out$) the interaction, integrated over the pulse duration $T$ :
$\mathbf{S}^{in/out}=\int_0^T dt s\left(0/L,t\right) $, and the
collective atomic operators \textit{before}/\textit{after} the
interaction $\mathbf{A}^{in/out}=\int_0^L dz \hat
A\left(z,0/T\right)$. $s$ and $\hat A$ have been normalised so that
$\mathbf{S}^{in/out}$ and $\mathbf{A}^{in/out}$ are dimensionless.
We note that the evolution equations~(\ref{evoat},\ref{inoutphot})
can alternatively be deduced following the methods
of~\cite{cohen1,cohen2} for the atoms and~\cite{laloe1,laloe2} for
the photons.

\section{Vectorial Hamiltonian}\label{vectoriel}

For $\alpha_T=0$, the Hamiltonian (\ref{eq:Hint}) reduces to the well-known ``QND" Hamiltonian
$\hbar \frac{\sigma\Gamma}{4A\Delta}\frac{\alpha_V}{2} S_z F_z$, which allows for non destructively
measuring $F_z$ via a measurement of the conjugate observable of $S_z$, as was shown in
\cite{kuzmichprl,polzikX,polzikint,mabusci}. We briefly review the principle of this conditional
squeezing of the orientation before generalizing it to an alignment in the next section.

Prior to the measurement of $\mathbf{F}_z$, the atoms are prepared
in a coherent spin state oriented along $x$, i.e. the atoms are
pumped into an eigenstate of $\mathbf{F}_x$. The values of the
components orthogonal to the mean spin, $\mathbf{F}_y$ and
$\mathbf{F}_z$, are unknown \textit{a priori}, and because of the
commutation relation $\left[ \mathbf{F}_y,\; \mathbf{F}_z \right] =
i \mathbf{F}_x = i N$, their standard deviations satisfy $\Delta
\mathbf{F}_y \Delta \mathbf{F}_z \geq N /2 $. When there exist no
correlation between the transverse components, such as in a sample
prepared by optical pumping, $\Delta \mathbf{F}_y = \Delta
\mathbf{F}_z = \sqrt{N/2}$. The atoms are placed in zero-magnetic
field. The probe is linearly polarized ($\langle
\vec{\mathbf{S}}\rangle=n\vec{x}$). Integrating the evolution
equations, one obtains the following input-output relations : \beqr
x^{out}&=&x^{in}+\kappa_V
s_z^{in}\\ p^{out}&=&p^{in}\\ s_y^{out}&=&s_y^{in}+\kappa_V p^{in}\\
s_z^{out}&=&s_z^{in}\eeqr The operators have been normalized so as
to have unity variance when they are in coherent states
($x,p=\mathbf{F}_{y,z}/\sqrt{N/2}$ and
$s_{y,z}=\textbf{S}_{y,z}/\sqrt{n}$). It is clear that by measuring
the fluctuations of $s_y^{out}$ one acquires information about the
fluctuations of $p$ ($F_z$ is measured non-destructively via the
Faraday rotation of the probe polarisation it induces). The
measurement is all the more accurate that the \textit{vectorial}
coupling strength
\beqr\label{kappaV}\kappa_V=\alpha_V\frac{\sigma\Gamma}{4A\Delta}\sqrt{\frac{Nn}{2}}\eeqr
is large. One therefore conditionally squeezes the atomic
orientation. The variances of the transverse components after the
measurement-induced projection of $s_y^{out}$ can easily be shown to
be those of a minimal spin-squeezed state
\cite{QNDgrangier,jfr,theseJuls} \beqr
V[x^{out}|s_y^{out}]=1+\kappa_V^2,\hspace{0.3cm}V[p^{out}|s_y^{out}]=\frac{1}{1+\kappa_V^2}
\eeqr

\section{Tensorial Hamiltonian}\label{tensoriel}

\begin{figure}
 \includegraphics[width=8cm]{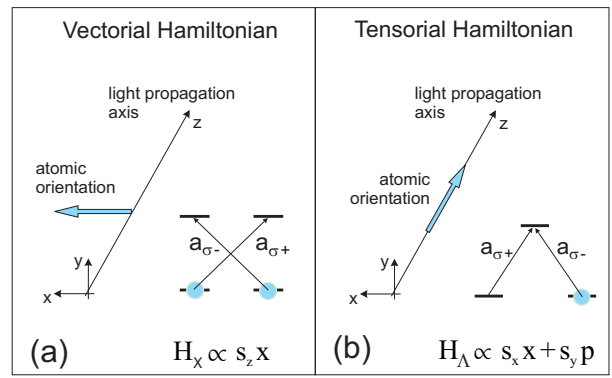}\\
  \caption{(a) 4-level scheme leading to a vectorial effective Hamiltonian $H_X\propto s_z x$.
  (b) 3-level $\Lambda$ scheme leading to a tensorial effective Hamiltonian $H_{\Lambda}\propto s_x x + s_y p$.}\label{LambdaX}
\end{figure}

\subsection{Single-pass interaction}\label{singlepass}
Another interesting situation is the opposite case of a purely
tensorial Hamiltonian, in which $\alpha_V=0$. In practice, the
interaction involves several hyperfine excited states $F'$, so that
it is possible to choose the detuning such that the various
vectorial contributions vanish $\sum_{F'}
\sigma_{F'}\alpha_V^{F'}\frac{\Gamma}{\Delta_{F'}}\simeq 0$, while
the total tensorial contribution $\sum_{F'} \sigma_{F'}
\alpha_T^{F'}\frac{\Gamma}{\Delta_{F'}}$ does not. It is then
possible to realize a conditional measurement of the alignment in
this particular situation. Let us assume that the atoms are prepared
in a coherent spin state along $z$. The conjugate transverse
components in this case are
$\frac{\mathbf{T}^2_2+\mathbf{T}^2_{-2}}{\sqrt{2}}$ and
$\frac{\mathbf{T}^2_2-\mathbf{T}^2_{-2}}{i \sqrt{2}}$, since
$[\frac{\mathbf{T}^2_2+\mathbf{T}^2_{-2}}{\sqrt{2}},\frac{\mathbf{T}^2_2-\mathbf{T}^2_{-2}}{i
\sqrt{2}}] = i N $. We normalize them as previously:
$x=(\mathbf{T}_2^2+\mathbf{T}_{-2}^2)/\sqrt{N}$ and
$p=(\mathbf{T}_2^2-\mathbf{T}_{-2}^2)/(i \sqrt{N})$ and assume a
circularly polarized probe: $\langle
\vec{\mathbf{S}}\rangle=n\vec{z}$.

The result of the integration of Eqs.~(\ref{evoat},\ref{inoutphot})
can be found in Ref.~\cite{polziktensbis} and is reminded in
Appendix \ref{integration}. It yields input-output relationships
involving complex spatiotemporal modes for the fields and the atoms.
For a thin medium ($\kappa_T \ll 1$), they lead to the following
input-output relations :
\beqr\label{sg1}
x^{out} &=& x^{in} + \kappa_T s_y^{in}\\
p^{out}&=& p^{in} - \kappa_T s_x^{in}\\
s_x^{out}&=&s_x^{in} + \kappa_T p^{in}\\
s_y^{out}&=&s_y^{in} - \kappa_T x^{in}\label{sg4}\eeqr with a
\textit{tensorial} coupling strength given by \beqr\label{kappaT}
\kappa_T=\alpha_T\frac{\sigma\Gamma}{8A\Delta}\sqrt{Nn}\eeqr
($\sum_{F'}
\alpha_T^{F'}\frac{\sigma_{F'}\Gamma}{8A\Delta_{F'}}\sqrt{Nn}$ if
several excited states are involved).

The interaction is obviously not QND, since both components of the
spin are now modified by the field, and conversely. This arises from
the fact that the effective Hamiltonian in this case, $H_{\Lambda}
\propto s_x x + s_y p$, is quite different of the previous vectorial
situation $H_{X} \propto s_z x$, and now involves both quadratures
(Fig.~\ref{LambdaX}). As noted in \cite{dantanEIT}, this tensorial
Hamiltonian corresponds to a linear coupling between two harmonic
oscillators which, when resonant, allows for efficient quantum state
transfer between atomic and light variables and may be used in
quantum memory protocols. As the coupling strength $\kappa_T$ is
increased, $x^{out}$ and $s_y^{out}$ (and $p^{out}$ and $s_x^{out}$,
respectively) coherently exchange their fluctuations, and it can
indeed be shown that, when the collective coupling strength
$\kappa_T$ is large, the field fluctuations are efficiently mapped
onto the atoms and vice-versa \cite{dantanPRA05,polziktensbis}.

However, since the atomic variables evolve during a single-pass
``tensorial" interaction, it is a priori not well-suited for QND
measurements. Nevertheless, it is still possible to perform a
conditional measurement of the alignment by using two ensembles $a$
and $b$ (or by making two successive passes in one ensemble) with
opposite mean orientations, such that $H_{int}\propto
(x_a+x_b)s_x+(p_a+p_b)s_y$ and $[x_a+x_b,\;p_a+p_b]=0$. As will be
detailed in the next Sections, this restores the QND character of
the interaction. The physical interpretation is that both field
quadratures are written onto the atoms in each ensemble, but,
because of the opposite orientations, their contributions cancel
out, leaving the total alignment components unchanged, while the
field still carries out information about both atomic alignment
components.

\subsection{Double-pass interaction}\label{doublepass}
\begin{figure}
  \includegraphics[width=8cm]{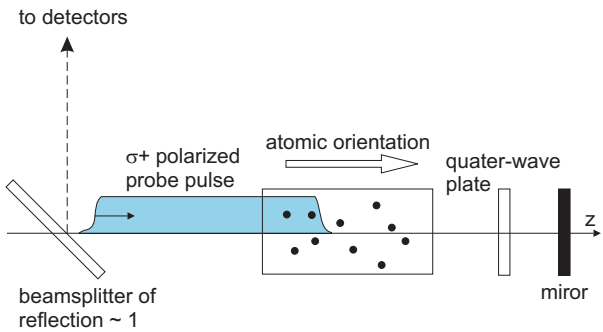}\\
  \caption{Schematic of the double pass configuration proposed to perform a QND
  measurement of a collective atomic alignment.}\label{aller_retour}
\end{figure}

\begin{figure}
  \includegraphics[width=8cm]{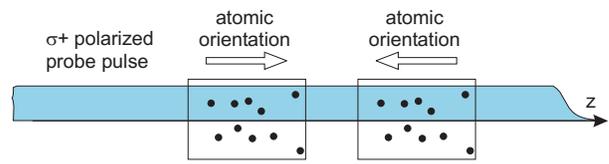}\\
  \caption{Schematic of the double ensemble configuration proposed to perform a QND
  measurement of a collective atomic alignment.}\label{double}
\end{figure}

\label{doublepassec}We first consider the double-pass geometry
depicted on Fig.~\ref{aller_retour}. A quarter-wave plate is
inserted between the cell and the mirror, its neutral axis being
aligned along $x$. We assume that the same pulse
\textit{successively} propagates back and forth in the atomic
ensemble, with no temporal overlap. This is different from the
situation of Refs.~\cite{takeushi,doublepulse}, where the pulse
interacts with itself in the atomic medium, so that the non linear
coupling allows for unconditional squeezing. We also note that
similar ideas have been proposed for quantum memories and squeezing
generation in Ref.~\cite{muschik} in the case of a purely vectorial
hamiltonian. In~\cite{muschik}, the second pass is used to couple
the second quadrature of light to the atoms. Again, the tensorial
situation is quite different, since the Hamiltonian directly couples
the two quadratures of light to the atoms. However, due to the $ s_x
x + s_y p$ form of the Hamiltonian, in order to perform a QND
measurement of the alignment, one has to compensate for extra
precession terms, which can be successfully done with a double-pass.

The double-pass interaction subsequently leads to \beqr
x^{out'}&=& x^{in}\label{dpe1}\\
p^{out'}&=& - p^{in} + 2 \kappa_T s_x^{in}\\
s_x^{out'} &=& s_x^{in}\\
s_y^{out'} &=& s_y^{in} - 2 \kappa_T x^{in}\label{dpe2}\eeqr The
measurement of $s_y^{out'}$ (resp. $s_x^{out'}$) projects $x^{out'}$
(resp. $p^{out'}$) in a state with reduced variance
$V[x^{out'}|s_y^{out'}] = V[p^{out'}|s_x^{out'}] = 1 - 4 \kappa_T^2
+ o(\kappa_T^2)$. For a moderate value of $\kappa_T=0.35$, these
variances are $\sim 0.5$, significantly smaller than the standard
quantum limit. A rigorous derivation of the conditional variance,
including the terms of order $\kappa_T^2$ in the input-output
relations (\ref{sg1}-\ref{sg4}), can be obtained from the exact
results of (\ref{int1}-\ref{int4}) and leads to exactly the same
conditional variance. The measurement of $x$ performed this way is
fully QND only for small values of $\kappa_T$, and will only result
in a limited squeezing in principle. We now turn to a situation
allowing for a QND measurement of the alignment for any value
$\kappa_T$.

\subsection{Double-cell interaction}\label{doublecell}

Alternatively, a single-pass interaction can be performed with two
atomic cells having opposite orientations - as in \cite{polzikint} -
in order to entangle the alignment components of two atomic
ensembles. As shown in Fig.~\ref{double}b the light pulse propagates
through two ensembles (a) and (b) prepared with opposite orientation
$\langle \mathbf{F}_z^a\rangle=-\langle \mathbf{F}_z^b\rangle=N$, so
that the input-output relationships now read \beqr
\left( x_a + x_b \right)^{out} &=& \left( x_a + x_b \right)^{in}\label{eq:twocellx}\\
\left( p_a + p_b \right)^{out} &=& \left( p_a + p_b \right)^{in}\\
s_x^{out}&=&s_x^{in}+\kappa_T \left( p_a + p_b \right)^{in} \label{eq:twocellsx}\\
s_y^{out}&=&s_y^{in}-\kappa_T \left( x_a + x_b \right)^{in}\label{eq:twocellsy}\eeqr If the probe
pulse duration is much longer than the time required to propagate through the two cells, the pulse
interacts \textit{simultaneously} with the two ensembles. In this experimentally accessible
situation, the previous relations hold to any order in $\kappa_T$, and the measurement is perfectly
QND. Similarly to the vectorial situation, measuring $s_y^{out}$ squeezes the variance of $x_a +
x_b$ to $2/\left(1+2 \kappa_T^2\right)$. Note that one has $\left[ x_a + x_b, p_a + p_b \right] = i
\left(F_{z \, a} + F_{z\, b}\right)2/N = 0$, since the two ensembles have opposite orientations. It
is therefore possible to squeeze not only the fluctuations of $x_a + x_b$, but also those of $p_a +
p_b$. As can be seen from Eq.~(\ref{eq:twocellsx}), sending a second pulse and detecting
$s_x^{out}$ instead of $s_y^{out}$ allows for squeezing $p_a +p_b$, leaving the alignment of the
two ensembles entangled. The expected value of entanglement obtained is $\Delta_{EPR} = \Delta^2
\left( x_a + x_b \right) + \Delta^2 \left( p_a + p_b \right)= 4/\left(1+2\kappa_T^2\right)< 4$.
Note that the result is the same as in the vectorial situation of \cite{polzikint}, but the
physical situation is rather different, since the tensorial situation requires a double-pass for
the alignment measurement to be completely QND.

\section{Atomic noise and experimental values for $^{87}$Rb}\label{noise}

\subsection{General Hamiltonian and non-zero frequency noise measurements}

For a single pass and in the case of a non-zero $\kappa_V$, the
input-output relations read to first order in $\kappa_V, \kappa_T$
\beqr\label{sgv1}
x^{out} &=& x^{in} + \kappa_T s_y^{in} - \kappa_V \sqrt{2n/N} p^{in}\\
p^{out}&=& p^{in} - \kappa_T s_x^{in} + \kappa_V \sqrt{2n/N} x^{in}\\
s_x^{out}&=&s_x^{in} + \kappa_T p^{in} - \kappa_V \sqrt{2N/n} s_y^{in}\\
s_y^{out}&=&s_y^{in} - \kappa_T x^{in} + \kappa_V \sqrt{2N/n}
s_x^{in}\label{sgv4}\eeqr In the double-pass geometry described in
Sec.~\ref{doublepassec}, the vectorial contributions cancel out and
Eqs.~(\ref{dpe1}-\ref{dpe2}) are left unchanged, so that the
alignment can still be conditionally squeezed in this scheme. In the
double-cell configuration, the vectorial contributions to the field
evolution (Faraday rotation) naturally cancel out, but the vectorial
contributions to the atom evolution (light-shifts) do not. However,
a $z$-aligned magnetic field with Larmor frequency $\Omega_L = -
\frac{\sigma \Gamma}{8 A \Delta} \alpha_V S_z$ can compensate for
these light-shifts.

Another experimentally relevant issue is the measurement of the
Stokes parameters fluctuations. Technical noise is in general
smaller than the quantum fluctuations of light only for
higher-frequency components (typically above 0.1-1 MHz). It is
therefore important to consider whether the schemes proposed in
Sec.~\ref{doublepass} and Sec.~\ref{doublecell} can be extended to
non-zero frequency noise measurements. In the double-cell
configuration, it can easily be done by means of a $z$-aligned
magnetic field. The Larmor precession couples $x$ and $p$, but in
the frame rotating at $2 \Omega$ ($\Omega$ is defined by $\hbar
\Omega = \mu B_z$, where $\mu$ is the magnetic moment of the ground
level and $B_z$ the magnetic field value). The input-output
relations (\ref{eq:twocellx}-\ref{eq:twocellsy}) then remain
unchanged when making the substitution $x\rightarrow (x)_{2\Omega} =
x\;\cos(2 \Omega t)+p\; \sin(2 \Omega t),$ $p\rightarrow
(p)_{2\Omega} = p\;\cos(2 \Omega t) - x\; \sin(2 \Omega t),$ etc. It
is thus possible to measure in a QND manner the atomic operators
$x_{2 \Omega}$ and $p_{2 \Omega}$ through their imprints on the
sidebands components of $S_x$ or $S_y$. Unfortunately, this
technique cannot be used in the double-pass configuration, since the
magnetic field would have to be reversed between the first and the
second pass, which is not very realistic experimentally. However, if
measurements of the Stokes operators at non-zero frequency are more
easily shot-noise limited, it was shown in \cite{mabusci} that
strong spin-squeezing could still be obtained experimentally in a
zero-magnetic field/frequency situation.

\subsection{Atomic noise considerations}

We now discuss the intrinsic limitations brought by spontaneous
emission noise in the tensorial Hamiltonian case. For the sake of
simplicity we study the case of a $1\rightarrow 0, 1$ transitions,
with atoms oriented along $z$. Using the Heisenberg-Langevin
evolution equations and the quantum regression theorem, we obtain
for a $1\rightarrow 0$ transition (for which $\alpha_V=-1/2$,
$\alpha_T=-1$ and $\kappa_T=\kappa_V \sqrt{2} \equiv\kappa_0$):
\begin{eqnarray}
x^{out} &=& x^{in} \sqrt{1- \varepsilon_{a}} +
\sqrt{\varepsilon_{a}} f_x \nonumber \\ && + \kappa_0
s_y^{in}-\kappa_0 \sqrt{\frac{n}{N}}p^{in} -
\frac{\gamma}{\Delta\sqrt{2}}\kappa_0 s_x^{in}\\
p^{out} &=& p^{in} \sqrt{1- \varepsilon_{a}} +
\sqrt{\varepsilon_{a}} f_p  \nonumber \\ && - \kappa_0
s_x^{in}+\kappa_0 \sqrt{\frac{n}{N}}x^{in} -
\frac{\gamma}{\Delta\sqrt{2}}\kappa_0 s_y^{in}\\
s_x^{out} &=& s_x^{in} \sqrt{1- \varepsilon_{p}} +
\sqrt{\varepsilon_{p}} f_{s_x}  \nonumber \\ && + \kappa_0
p^{in}-\kappa_0 \sqrt{\frac{n}{N}}s_y^{in} -
\frac{\gamma}{\Delta}\kappa_0 x^{in}\\
s_y^{out} &=& s_y^{in} \sqrt{1- \varepsilon_{p}} +
\sqrt{\varepsilon_{p}} f_{s_y}  \nonumber \\ && - \kappa_0
x^{in}+\kappa_0 \sqrt{\frac{n}{N}}s_x^{in} -
\frac{\gamma}{\Delta}\kappa_0 p^{in}
\end{eqnarray}
with $\varepsilon_{a} = - \frac{\kappa_0
\Gamma}{\Delta}\sqrt{\frac{n}{N}}$, $\varepsilon_{p} = - \frac{
\kappa_0 \Gamma}{\Delta}\sqrt{\frac{N}{n}}$ and $f_{x,p}$,
$f_{s_x,s_y}$ standard vacuum noise operators with variance unity.

For the $1\rightarrow 1$ transition, one has $\alpha_V=-3/4$ and
$\alpha_T=3/2$, so that $\kappa_T= - \kappa_V \sqrt{2}
\equiv\kappa_1$ and similar equations can be derived. Choosing the
detunings such that $\kappa_0 = \kappa_1\equiv\kappa/2$ cancels the
vectorial terms finally yields the following input-output
relationships
\begin{eqnarray}\nonumber x^{out} &=& x^{in} \sqrt{1-
\varepsilon_{a}} + \sqrt{\varepsilon_{a}} f_x + \kappa s_y^{in}+
\frac{\varepsilon'}{\sqrt{2}} s_x^{in}\\\nonumber p^{out} &=& p^{in}
\sqrt{1- \varepsilon_{a}} + \sqrt{\varepsilon_{a}} f_p - \kappa
s_x^{in}+ \frac{\varepsilon'}{\sqrt{2}} s_y^{in}\\\nonumber
s_x^{out} &=& s_x^{in} \sqrt{1- \varepsilon_{p}} +
\sqrt{\varepsilon_{p}} f_{s_x} + \kappa p^{in} + \varepsilon'
x^{in}\\\nonumber s_y^{out} &=& s_y^{in} \sqrt{1- \varepsilon_{p}} +
\sqrt{\varepsilon_{p}} f_{s_y} - \kappa x^{in} + \varepsilon' p^{in}
\end{eqnarray} with $\varepsilon_{a} = \frac{\kappa \Gamma}{2}\sqrt{\frac{n}{N}}\left(\frac{1}{\Delta_1}
-\frac{1}{\Delta_0}\right),\hspace{0.1cm} \varepsilon_{p} =
\frac{\kappa \Gamma}{2} \sqrt{\frac{N}{n}}\left(\frac{1}{\Delta_1}
-\frac{1}{\Delta_0}\right)$ and $\varepsilon'=- \kappa
\frac{\Gamma}{4} \left(\frac{1}{\Delta_1}
+\frac{1}{\Delta_0}\right)$. One retrieves beamsplitter-like
relations for the losses, similar to those of \cite{polzikX}.
$\varepsilon_{p}$ simply describes absorption of the probe caused by
spontaneous emission : the probe field is damped by a factor
$\sqrt{1-\varepsilon_{p}}$, and some uncorrelated vacuum noise
$\sqrt{\varepsilon_{p}} f_{s_x,s_y}$ is consequently added, as for
the propagation through a beamsplitter with transmission
$\sqrt{1-\varepsilon_{p}}$. $\varepsilon_{a}$ describes the
symmetrical process for the atoms : the probe, because of
spontaneous emission, induces optical pumping towards a $z$-aligned
coherent spin state (which is similar to mixing the probe with some
vacuum). A difference with the vectorial situation is the presence
of small contamination terms $\varpropto \varepsilon'$. They also
correspond to optical pumping processes which tend to align $x$ and
$p$ along $s_x$ and $s_y$. To minimize the effect of spontaneous
emission noise, one has to choose $n\sim N$ in order to have
$\varepsilon_a\sim\varepsilon_p$, as in Ref.~\cite{polzikX}.
Finally, the total spontaneous emission contribution in the
double-pass or double-cell configurations is finally obtained by
doubling $\varepsilon_{a}$, $\varepsilon_{p}$ and $\varepsilon'$ in
the above equations.

\subsection{Experimental values for $^{87}$Rb}\label{discussion}

Based on these considerations we discuss the values of squeezing or
entanglement that can be expected in experiments with $^{87}$Rb. We
assume an interaction on the $D_2$ line with the atoms in the $F=1$
ground state. For room temperature vapor cells, taking into account
the Doppler broadening, no detuning allows for completely canceling
$\kappa_V$. On the contrary, for cold atoms with negligible Doppler
broadening, $\kappa_V$ can be canceled for a probe laser
blue-detuned by $\Delta_0=\mathrm{38}$ MHz from the $F'=0$ excited
level (red-detuned by 34 MHz from $F'=1$ and 191 MHz from $F'=2$).
For typical values for the density and volume ($10^{11}$ cm$^{-3}$
and $0.5$ mm$^3$) of a cold atom cloud produced using a
magneto-optical trap (the latter being switched off during the
measurement), leading to $N=0.5\times 10^{8}$ and taking a pulse of
intensity 1 $\mu$W and duration $0.5$~$\mu$s containing $n\sim
0.5\times 10^{8}$ photons (saturation parameter $\sim 10^{-3}$,
considering a cross-section $A=1$ mm$^2$), the previous calculations
predict $\kappa_T\sim - 0.42$ and $\sim -5dB$ of squeezing in the
quantum fluctuations of $\mathbf{T}^2_2+\mathbf{T}^2_{-2}$ or
$\mathbf{T}^2_2-\mathbf{T}^2_{-2}$. Higher values of $\kappa_T$ (and
hence higher squeezing values) can be reached for longer probe
pulses, provided that the duration of the pulses remains smaller
than the relaxation time of the Zeeman coherence, or by the use of a
dipole trap to increase the optical depth \cite{deechaniz}. For
these parameters, in a double-pass or double-cell configuration,
$\varepsilon_a = \varepsilon_p = 0.14$ and $\varepsilon' \lesssim
10^{-4}$ (the contribution of the $F'=2$ level is $\sim 30$ times
smaller than those of the $F'=0,1$ levels, and is not considered).
As the fluctuations are predicted here to be reduced by a factor
smaller than $\sim 1/0.14$, the noise added by spontaneous emission
can be neglected.

For the sake of comparison, we now discuss the relative strengths of
tensorial and vectorial conditional measurements. In atomic vapors
close to room temperatures, the detuning is usually chosen bigger
than the doppler broadening in order to avoid
absorption~\cite{polzikint}. It implies that the detuning has to be
large as compared to the hyperfine structure, and, since one has
$\sum_{F'} \sigma_{F'} \alpha_T^{F'} = 0$ for alkali atoms, it means
that $\kappa_T \propto \sum_{F'} \sigma_{F'} \alpha_T^{F'} /
\Delta_F' \sim 0$, i.e. the effective Hamiltonian is then almost
purely vectorial. This situation is obviously much more favorable
for orientation than for alignment squeezing.

\begin{figure}
  \includegraphics[width=8cm]{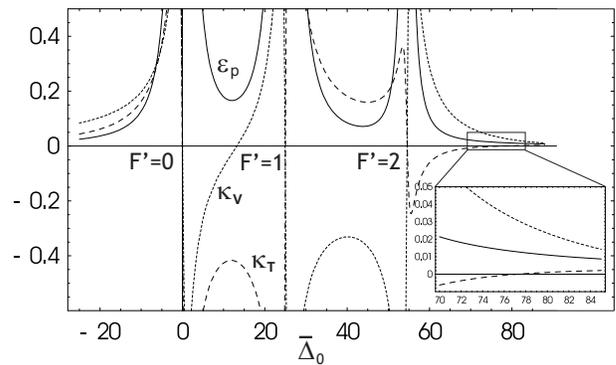}\\
  \caption{Vectorial and tensorial coupling strength $\kappa_V$ (dotted) and $\kappa_T$ (dashed),
  and amplitude of the noise added to the probe $\varepsilon_p$ (plain)
  as functions of the normalised detuning $\overline{\Delta_0}=\Delta_0/(\Gamma/2)$
  between the probe and the $F=1 \rightarrow F'=0$ transition of $^{87}$Rb $D_2$ line.
  The experimental parameters are detailed in Sec.\ref{discussion}. The insert zooms on the detuning area
where the Hamiltonian is purely vectorial.} \label{fig:kappas}
\end{figure}

However, for a doppler-free medium, it is possible to reduce the
detuning while maintaining a small absorption. In this case, as can
be seen from Fig.~\ref{fig:kappas}, both $\kappa_T$ and $\kappa_V$
(and hence alignment and orientation squeezing) may have similar
values. To compare these values, we consider the case of a purely
tensorial Hamiltonian (i.e $\kappa_V=0$, obtained for
$\Delta_0=\mathrm{38}$ MHz, $\overline{\Delta_0}=13.2$), and the
case of a purely vectorial one (i.e $\kappa_T=0$, obtained for
$\Delta_0=\mathrm{222}$ MHz, $\overline{\Delta_0}=77$). In the first
one, for the experimental parameters given above, $\kappa_T=-0.42$,
$\kappa_V=\mathrm{0}$ and $\varepsilon_p=\varepsilon_a=0.14$ whereas
in the second , $\kappa_T=\mathrm{0}$ , $\kappa_V=0.03$ and
$\varepsilon_p=\varepsilon_a=0.01$. This shows that the common idea
that the vectorial coupling strength is bigger than the tensorial
one is not necessarily true for cold atom samples when the hyperfine
structure is taken in account.

\section{Conclusion}
We have shown how to perform a QND measurement of a collective
atomic alignment. This extends the possibility to manipulate
high-angular momentum components of a collective spin beyond the
vectorial Hamiltonian interaction commonly used so far in
experiments \cite{kuzmichprl,polzikint,mabusci}. Noticeable physical
differences are found between the purely vectorial Hamiltonian
situation and the tensorial situation. In particular, if it had been
noted in previous work \cite{dantanEIT,polziktensbis} that the
tensorial situation may lead to coherent atom-field quantum state
transfer and storage, we have shown here that it also allows for
performing a QND measurement of the atomic alignment, provided that
two ensembles or two successive passes are used. Substantial
conditional squeezing values are still predicted for realistic
experimental situations with cold atomic samples. We also note that
these measurements can be used to continuously control the atomic
spin fluctuations via feedback \cite{mabusci}. The different
feedback mechanisms that may be used to squeeze an atomic alignment
will be presented elsewhere.

\begin{acknowledgments}
 The authors would like to thank W. Gawlik for fruitful discussions
 on precision atomic magnetometry.
\end{acknowledgments}

\appendix
\section{Polarizability}\label{alpha}

The polarisabilities $\alpha_V^{F'}$ and $\alpha_T^{F'}$ are given
by
\begin{widetext}\begin{eqnarray*} \alpha_V^{F'} &=&
\frac{3(2J'+1)}{2(2F'+1)(2J+1)}\left(
    -\frac{2F-1}{F}\delta^{F'}_{F-1}
    -\frac{2F+1}{F(F+1)}\delta^{F'}_{F}
    +\frac{2F+3}{F+1}\delta^{F'}_{F+1}
\right)\\
\alpha_T^{F'} &=& - \frac{3(F+1)(2J'+1)}{2(2F'+1)(2J+1)}\left(
    \frac{1}{F}\delta^{F'}_{F-1}
    -\frac{2F+1}{F(F+1)}\delta^{F'}_{F}
    +\frac{1}{F+1}\delta^{F'}_{F+1}
\right)
\end{eqnarray*} where $\delta^F_{F'}$ is Kronecker's symbol. The resonant cross-section between two levels with an isotropically populated
ground state is
$$
\sigma_{F'} = \sigma_{2level}\frac{2(2J+1)(2F'+1)}{3} \left\{
\begin{array}{ccc}
                   J' & 1 & J \\
                   F & I & F' \\
                 \end{array}
\right\}^2
$$
with $\sigma_{2level} = \frac{3\lambda^2}{2\pi}$. The commutators
between the irreducible tensorial operators $T^k_q$
are~\cite{omont}:
\begin{eqnarray*} \left[ T^{k_1}_{q_1}(F_g),
T^{k_2}_{q_2}(F_{g})\right] =
    \sum_{K, Q} (-1)^{K+2F_g}
    \sqrt{(2k_1+1)(2k_2+1)}
    \left\{
    \begin{array}{ccc}
                   k_1 & k_2 & K \\
                   F_g & F_g & F_g \\
                 \end{array}
    \right\} \times \\
    \left\langle\ k_1 k_2 q_1 q_2 , K Q
    \right\rangle
    (1-(-1)^{k_1+k_2+K}) T^K_Q(F_g)
\end{eqnarray*}\end{widetext}

\section{Tensorial situation : solutions of the evolution equations~(\ref{evoat}-\ref{inoutphot})}\label{integration}

We assume a single-pass interaction with $\alpha_V=0$, as in
Sec.~\ref{singlepass}. After changing the spatiotemporal frame
$(z,t)\rightarrow (z=z, t=t-z/c)$ and making the system
dimensionless $(z,t)\rightarrow (\mathbf{z} = z/L, \mathbf{t}=t/T)$,
the integration of Eqs.~(\ref{evoat}-\ref{inoutphot}) yields
\cite{polziktensbis}
\beqr\nonumber x^{out}&=&x^{in}-\kappa_T\int_0^1d\mathbf{z}\int_0^\mathbf{z}d\mathbf{z'}
x^{in}(\mathbf{z'})\frac{J_1(2\kappa_T\sqrt{\mathbf{z-z'}})}{\sqrt{\mathbf{z-z'}}}\\\nonumber
&&+\kappa_T\int_0^1d\mathbf{t} s_y^{in}(\mathbf{t})\left(\int_0^1
d\mathbf{z}
J_0(2\kappa_T\sqrt{\mathbf{z}(1-\mathbf{t})})\right)\\\label{int1}\\\nonumber
p^{out}&=&p^{in}-\kappa_T\int_0^1d\mathbf{z}\int_0^\mathbf{z}d\mathbf{z'}
p^{in}(\mathbf{z'})\frac{J_1(2\kappa_T\sqrt{\mathbf{z-z'}})}{\sqrt{\mathbf{z-z'}}}\\\nonumber\label{int2}
&&-\kappa_T\int_0^1d\mathbf{t} s_x^{in}(\mathbf{t})\left(\int_0^1
d\mathbf{z}
J_0(2\kappa_T\sqrt{\mathbf{z}(1-\mathbf{t})})\right)\\\label{int2}\eeqr
and symmetrical equations for the fields \beqr\nonumber
s_x^{out}&=&s_x^{in}-\kappa_T\int_0^1 d\mathbf{t}\int_0^\mathbf{t}
d\mathbf{t'}s_x^{in}(\mathbf{t'})\frac{J_1(2\kappa_T\sqrt{\mathbf{t-t'}})}{\sqrt{\mathbf{t-t'}}}\\\nonumber
&&+\kappa_T\int_0^1 d\mathbf{z} p^{in}(\mathbf{z})\left(\int_0^1
d\mathbf{t} J_0(2\kappa_T
\sqrt{\mathbf{t}(1-\mathbf{z})})\right)\\\label{int3}\\\nonumber
s_y^{out}&=&s_y^{in}-\kappa_T\int_0^1 d\mathbf{t}\int_0^\mathbf{t}
d\mathbf{t'}s_y^{in}(\mathbf{t'})\frac{J_1(2\kappa_T\sqrt{\mathbf{t-t'}})}{\sqrt{\mathbf{t-t'}}}\\\nonumber
&&-\kappa_T\int_0^1 d\mathbf{z} x^{in}(\mathbf{z})\left(\int_0^1
d\mathbf{t} J_0(2\kappa_T
\sqrt{\mathbf{t}(1-\mathbf{z})})\right)\\\label{int4} \eeqr where
$J_0$ and $J_1$ are the standard first order Bessel functions and
the operators have been normalized so as to have unity variances
when in coherent states. At first order in $\kappa_T$, one retrieves
Eqs.~(\ref{sg1}-\ref{sg4}).

\bibliography{allign_sqz_cvik}
\end{document}